# A MIXTURE MODEL FOR UNSUPERVISED TAIL ESTIMATION

LARS HOLDEN AND OLA HAUG

Abstract. This paper proposes a new method to combine several densities such that each density dominates a separate part of a joint distribution. The method is fully unsupervised, i.e. the parameters in the densities and the thresholds are simultaneously estimated. The approach uses cdf functions in the mixing. This makes it easy to estimate parameters and the resulting density is smooth. Our method may be used both when the tails are heavier and lighter than the rest of the distribution. The presented model is compared with other published models and a very simple model using a univariate transformation.

Mixing functions; Heavy and Light Tailed Distributions; Maximum Likelihood; Mixture Models.

## 1. INTRODUCTION

In many applications, the tail of a probability distribution is of particular interest, e.g. prediction of floods or estimation of financial reserves in insurance. Because extreme data are rare, it is difficult to fit tail models and to support parametric model choices convincingly. Most papers study this problem in one dimension assuming a heavy tail. The approach presented in this paper mix different distributions that describes different parts of a new joint distribution. The joint distribution may have heavier or lighter tail(s) compared to the tail(s) of the distribution that is used in the central part of the joint distribution. Moreover, it is possible to generalise to $R^n$.

Common practice in extreme value modelling is to fix a threshold $u$ and to fit a distribution, e.g. a generalized Pareto Distribution (GPD), to the data exceeding $u$. There is a number of methods to estimate the parameters once $u$ is fixed, see for instance [16], [7] and references therein. As is well known, the estimates depend significantly on the choice of the threshold, see for instance [10], Figure 6.2.8. In order to reduce model bias, the threshold $u$ should be chosen large, but this often leaves very few data points for the estimation of the parameters. Hence, the resulting parameter estimates will have large variances. Moreover, the selection of an appropriate threshold is a difficult task in practice, see for instance [8], [17], [10], and [14]. Often a supervised analysis is performed, selectively and off-line as part of a monitoring scheme. For practitioners, who usually need to perform their data analyses regularly, it would be convenient to have automatic and robust approaches that do not require an a priori tuning of a threshold. Such an unsupervised approach to tail estimation would be of particular relevance in automatic real-time monitoring of financial, industrial and environmental quantities, for instance for warning purposes.

Recently, [9] and [11] have proposed two new ways of addressing this question. The paper [9] suggests a robust model validation mechanism to guide the threshold selection. The procedure assigns weights between zero and one to each data point, where a high weight means that the point should be retained since a GPD model is fitting it well. The author suggests to start with a low threshold $u$ and increase it, thus reducing the number of data points, until all data left have weights close to

1. This is a promising method, but thresholding is still needed at the level of the weights.

The paper [11] was the first to suggest a fully unsupervised approach to tail estimation. Their approach has three key ingredients. First, the model consists of two components, one representing the central part of the distribution and the other the tail. Second, all data are modelled in one mixture model, and finally, the parameters in the two distributions and the mixing parameters are simultaneously estimated.

Our method is based on the same ideas as those in [11], but we mix the cumulative distribution functions (cdfs) instead of densities. This makes our approach computationally more efficient, which in turn makes it easier to generalise to higher dimensions. Threshold estimation is more complicated in higher dimensions, since the threshold is a surface. However, in our approach, this is handled efficiently. This paper only shows examples in one dimension. The multivariate examples are reported separately, [12]. We also show how to use several different distributions for different parts of the tail(s). Also the method presented in [3] uses cdfs in the mixing. But our method gives a continuous density in contrast to the method presented in [3].

The proposed model is compared with a model based on a univariate transformation. The properties of the two models are quite similar. But when generalizing to several dimensions there are important differences. The model based on cdfs may combine different multivariate densities, but needs to calculate the cdfs and not only the densities in the evaluation. The model based on univariate transformation does not need to calculate the multivariate cdfs, but the properties are dominated by the properties of the chosen multivariate distribution. If we want to change the multivariate properties, the method may be combined with a copula approach.

In many applications it is needed to describe the entire distribution, not only the tail(s). In our methods, the user selects densities that he/she believes fits the different parts of the data. However, in some cases the density that is used for describing the tail also describes the central part of the distribution better than the density that is supposed to describe the central part of the density. Then the tail density may end up modelling most of the density leading to better overall match with data, but with poorer description of the tail. This is easy to identify from the estimated threshold and may for example be corrected by putting a prior on the threshold.

We first describe the cdf-model in Section 2 and the transformed model in Section 3. Section 4 compares our approach to other models. Then, in Section 5, the models are tested with synthetic data and financial data. Finally, Section 6 contains some concluding remarks.

## 2. The cdf-model

In this section we start with two one-dimensional components in the mixture, and then we show how the model may be generalised to several components.

2.1. **Two components.** Let $x \in R$ and let $G(x; \theta_G)$ and $F(x; \theta_F)$ be two cdfs that we want to combine to a cdf denoted $L$. Define a threshold $u$ and a mixing zone $(u - \varepsilon, u + \varepsilon)$ for $\varepsilon \geq 0$, and let the cdfs $G$ and $F$ determine the properties of $L$ below and above the mixing zone, respectively. Further, let both cdfs influence $L$ in the mixing zone. We will often mix truncated distributions. Hence, we define the truncated densities where

$$g_t(x; \theta_G) = \begin{cases} g(x; \theta_G) & \text{if } x < u \\ 0 & \text{if } x \geq u \end{cases}$$

and

$$f_t(x; \theta_F) = \begin{cases} 0 & \text{if } x < u \\ f(x; \theta_F) & \text{if } x \geq u. \end{cases}$$

The corresponding truncated cdfs are defined as

$$G_t(x; \theta_G) = \int_{-\infty}^{x} g_t(t; \theta_G) dt$$

and

$$F_t(x; \theta_F) = \int_{-\infty}^{x} F_t(t; \theta_F) dt$$

where we do not require that $G_t(\infty; \theta_G) = 1$ and $F_t(\infty; \theta_F) = 1$. We then define the mixed cdf by

$$L(x; \theta_L) = \kappa(G_t(q(x; \theta_q); \theta_G) + F_t(p(x; \theta_p); \theta_F)) \tag{1}$$

where $\kappa$ is defined such that $L(\infty; \theta_L) = 1$ and $q(x; \theta_q)$ and $p(x; \theta_q)$ are two monotone increasing mixing functions described below. Note that $\kappa$ is easily found by $\kappa = 1/(G_t(\infty; \theta_G) + F_t(\infty; \theta_F))$. The parameters of $L(x; \theta_L)$ include all the other parameters i.e. $\theta_L = (\theta_G, \theta_F, \theta_q, \theta_p)$. Equation (1) is a well-defined cdf when the truncated cdfs $G_t$ and $F_t$ and the mixing functions $q$ and $p$ satisfy the criteria specified below. The corresponding density $l(x; \theta_L)$ is given by

$$l(x; \theta_L) = \begin{cases} \kappa\, g(x; \theta_G) & \text{if } x < u - \varepsilon \\ \kappa\, (g(q(x; \theta_q); \theta_G)\, q'(x; \theta_q) + & \\ \qquad f(p(x; \theta_p); \theta_F)\, p'(x; \theta_p)) & \text{if } x \in (u - \varepsilon, u + \varepsilon) \\ \kappa\, f(x; \theta_F) & \text{if } x > u + \varepsilon \end{cases} \tag{2}$$

This requires that $q(x; \theta_q) = p(x; \theta_q) = x$ where it is applied for $x$ outside the mixing zone. The mixing functions $q$ and $p$ determine how $G$ and $F$ influence $L$ in the mixing zone. They are monotonously increasing functions defined on $R$ and with range equal to $R$. If we set $\varepsilon = 0$ and the two mixing functions equal to the identity function i.e. $q(x; \theta_q) = p(x; \theta_q) = x$, then we get the standard approach where only data above the threshold is used in the tail estimation and the joint distribution is discontinuous. In our approach, we want all data to be used in the estimation of a continuous density $l(x; \theta_L)$. Then we set $\varepsilon > 0$ and the function $q$ maps the interval $(-\infty, u + \varepsilon)$ onto $(-\infty, u)$ and $p$ maps the interval $(u - \varepsilon, \infty)$ onto $(u, \infty)$ as illustrated in Figure 1. In order to get the derivative of $l(x; \theta_L)$ to be continuous, we need to have $q'(u+\varepsilon; \theta_q) = q''(u+\varepsilon; \theta_q) = 0$ and $p'(u-\varepsilon; \theta_q) = p''(u-\varepsilon; \theta_q) = 0$. Further properties of the mixing function is discussed in Section 2.2. We have found that the two mixing functions $p(x; \theta_p)$ and $q(x; \theta_q)$ defined below work well. Define

$$q(x; \theta_q) = \begin{cases} x & x < u - \varepsilon \\ \frac{1}{2}(x + u - \varepsilon) + \frac{\varepsilon}{\pi} \cos(\frac{\pi(x-u)}{2\varepsilon}) & u - \varepsilon \leq x < u + \varepsilon \\ x - \varepsilon & u + \varepsilon \leq x, \end{cases} \tag{3}$$

$$p(x; \theta_p) = \begin{cases} x + \varepsilon & x < u - \varepsilon \\ \frac{1}{2}(x + u + \varepsilon) - \frac{\varepsilon}{\pi} \cos(\frac{\pi(x-u)}{2\varepsilon}) & u - \varepsilon \leq x < u + \varepsilon \\ x & u + \varepsilon \leq x. \end{cases} \tag{4}$$

where $\theta_q = \theta_p = (u, \varepsilon)$. Figure 2 shows an example where two Gaussian densities are mixed.

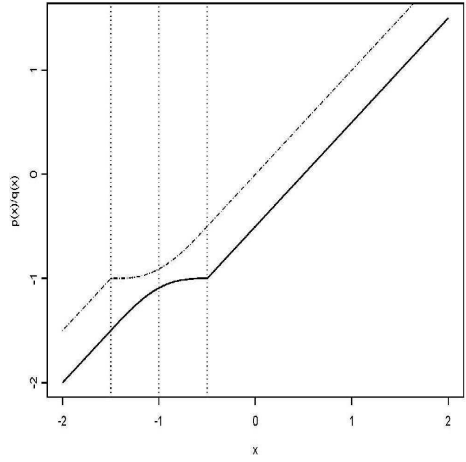

FIGURE 1. The mixing functions $p(x; \theta_p)$ (dotted line) and $q(x; \theta_q)$ (solid line) for $u = -1$ and $\varepsilon = 0.5$. The vertical bars correspond to $u - \varepsilon$, $u$ and $u + \varepsilon$, respectively.

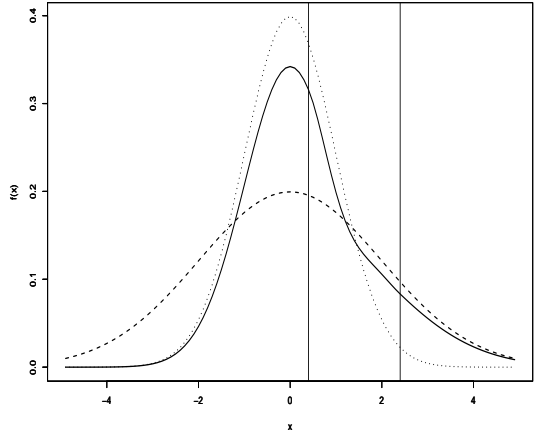

FIGURE 2. The two normal densities $g(x; \theta_G) \sim N(0, 1)$ and $f(x; \theta_F) \sim N(0, 4)$ are mixed with the mixing zone $(0.4, 2.4)$ The resulting mixture density $l(x; \theta_L)$ is given by the solid black line.

2.2. **Several components.** Equation (1) may easily be generalised to a mixture of several truncated cdfs $G_1, \ldots, G_k$ with parameters $\theta_{G_1}, \ldots, \theta_{G_k}$. Define a threshold $u_i$ and a mixing zone $(u_{i-1} - \varepsilon_{i-1}, u_i + \varepsilon_i)$, where $\varepsilon_i \geq 0$ and the truncated density $g_i(x; \theta_G) > 0$ only if $u_{i-1} < x < u_i$ for each component $i$. We assume that $u_0 = -\infty$ and $u_k = \infty$. The resulting cdf is given by

$$(5) \qquad L(x; \theta_L) = \kappa \sum_{i=1}^{k} G_i(q_i(x; \theta_{q_i}); \theta_{G_i}),$$

where $\kappa$ is defined such that $L(\infty; \theta_L) = 1$. Let $G_i(u_{i-1}; \theta_{G_i}) = 0$ and

$$G_i(x; \theta_{G_i}) = \int_{u_{i-1}}^{x} g_i(t; \theta_G) dt$$

for $i = 1, \cdots, k$.

The density $l(x; \theta_L)$ corresponding to the cdf $L(x; \theta_L)$ in Equation (5) is given by

(6)
$$
l(x; \theta_L) = \begin{cases} \kappa \, g_i(x; \theta_{G_i}) & \text{if } x < u_i - \varepsilon_i \\ \kappa \, (g_i(q_i(x; \theta_{q_i}); \theta_{G_i}) \, q'_i(x; \theta_{q_i}) + \\ \qquad g_{i+1}(q_{i+1}(x; \theta_{q_{i+1}}); \theta_{G_{i+1}}) \, q'_{i+1}(x; \theta_{q_{i+1}})) & \text{if } x \geq u_i - \varepsilon_i \end{cases}
$$

assuming we have $x \in (u_{i-1} + \varepsilon_{i-1}, u_i + \varepsilon_i)$ for a value of $i$. The first expression denotes the density between two consecutive mixing zones, and the other within a mixing zone.

Each mixing function $q_i$, with parameters $\theta_{q_i} = (u_{i-1}, \varepsilon_{i-1}, u_i, \varepsilon_i)$, maps the interval $(u_{i-1} - \varepsilon_{i-1}, u_i + \varepsilon_i)$ onto $(u_{i-1}, u_i)$. The mixing functions must be continuous and monotonously increasing. Moreover, in order to ensure a smooth transition between the densities, we should have

(7)
$$ q'_i(x; \theta_{q_i}) + q'_{i+1}(x; \theta_{q_{i+1}}) = 1, $$

in the mixing zone, and each $q_i$, $i = 1, \cdots, k$ should satisfy

(8)
$$ q'_i(u_{i-1} - \varepsilon_{i-1}; \theta_{q_i}) = 0, \qquad q'_i(u_{i-1} + \varepsilon_{i-1}; \theta_{q_i}) = 1, $$

(9)
$$ q'_i(u_i - \varepsilon_i; \theta_{q_i}) = 1 \quad \text{and} \quad q'_i(u_i + \varepsilon_i; \theta_{q_i}) = 0. $$

We avoid breakpoints in the density corresponding to the cdf $L$ in Equation (5) by also requiring

(10)
$$ q''_i(u_{i-1} - \varepsilon_{i-1}; \theta_{q_i}) = 0, \qquad q''_i(u_{i-1} + \varepsilon_{i-1}; \theta_{q_i}) = 0, $$

(11)
$$ q''_i(u_i - \varepsilon_i; \theta_{q_i}) = 0 \quad \text{and} \quad q''_i(u_i + \varepsilon_i; \theta_{q_i}) = 0. $$

One of the major problems in extreme value theory is to estimate the threshold $u$. We reduce this problem by defining the threshold $u_i$ from the equation

(12)
$$ g_i(u_i; \theta_{G_i}) = g_{i+1}(u_i; \theta_{G_{i+1}}). $$

If there are several values of $u_i$ that satisfies the equation, we may select the supremum or infinum of these values. Equation (12) ensures that there are not large changes in $l(x; \theta_L)$ in the mixing zones. Letting the threshold be a function of the other parameters instead of a separate parameter, reduces the number of parameters. In Section 5.1 we also show that this makes the estimation of the parameters in the model more stable. Equations (8) - (9) ensure that $l(x; \theta_L)$ has continuous derivative without requiring Equation (12). If $g_i, g_{i+1}, \cdots, g_k$ have heavier and heavier tails or lighter and lighter tails, Equation (12) is particularly natural. If the tails are heavier, then $\kappa$ is slightly less than 1 and if the tails are lighter, then $\kappa$ is slightly larger than 1.

There are several possible definitions for $q_i$ that satisfy the requirements given in Equations (7)-(9). We use

(13)
$$
q_i(x; \theta_{q_i}) = \begin{cases} x + \varepsilon_{i-1} & x < u_{i-1} - \varepsilon_{i-1} \\ \frac{1}{2}(x + u_{i-1} + \varepsilon_{i-1}) - \frac{\varepsilon_{i-1}}{\pi} \cos(\frac{\pi(x - u_{i-1})}{2\varepsilon_{i-1}}) & u_{i-1} - \varepsilon_{i-1} \leq x < u_{i-1} + \varepsilon_{i-1} \\ x & u_{i-1} + \varepsilon_{i-1} \leq x < u_i - \varepsilon_i \\ \frac{1}{2}(x + u_i - \varepsilon_i) + \frac{\varepsilon_i}{\pi} \cos(\frac{\pi(x - u_i)}{2\varepsilon_i}) & u_i - \varepsilon_i \leq x < u_i + \varepsilon_i \\ x - \varepsilon_i & u_i + \varepsilon_i \leq x. \end{cases}
$$

Figure 3 shows the $q_i$-function.

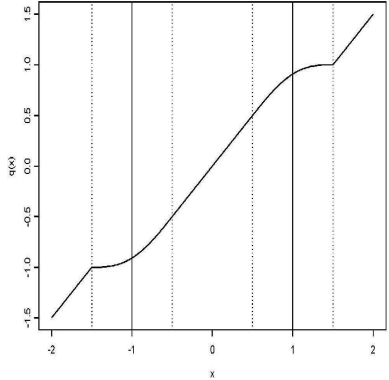

FIGURE 3. The transition function $q_i(x; \theta_{q_i})$ with $u_{i-1} = -1$, $u_i = 1$, and $\varepsilon_{i-1} = \varepsilon_i = 0.5$. The thresholds are given by vertical solid lines and the mixing zones are delimited by the vertical dotted lines. Note that the transition function maps the interval $(u_{i-1} - \varepsilon_{i-1}, u_i + \varepsilon_i)$ onto $(u_{i-1}, u_i)$.

## 3. THE TRANSFORMATION MODEL

An alternative to the model described in the previous section is to transform data to a known density like what is done in a normal score transform. We will present a method of this type where we focus on the tail behaviour and make it quite similar to the method presented in the previous section. Since we focus on the tails where there are few data points, we use a parametric transformation instead of an empiric transformation. The authors are well aware that the main argument for this model is that it is mathematically convenient and not that it is based on classical statistical principles. There are many similarities between this approach and the method described in the previous section, and the results are as good as for the other method.

Let $x \in R$ and let $G(x; \theta_G)$ be a cdf where we want to change the tail behaviour. Let $q(x; \theta_q)$ be a monotone increasing function and define the new cdf by the function

$$(14) \qquad L(x; \theta_L) = G(q(x; \theta_q); \theta_G)$$

which is a valid cdf. The density is obviously

$$(15) \qquad l(x; \theta_L) = g(q(x; \theta_q); \theta_G) q'(x; \theta_q)$$

where $g$ and $q'$ denote the derivative of $G$ and $q$ respectively. We get heavier tail if $|q(x; \theta_q)| < |x|$ and lighter tail if $|q(x; \theta_q)| > |x|$ for $x$ in the tail of $g$. There is a large variety of alternatives for the function $q$. Using the same notation as in the previous section we define a mixing zone $(u - \varepsilon, u + \varepsilon)$ where we let $q(x) = x$ in the central part of the distribution and $q(x) = f(x)$ on the tail side (outside) of the mixing zone. We have found that $f(x) = u(x/u)^\beta$ gives good results. If we let $G$ be the normal distribution, we see from Equation (15) that $l(x, \theta_L)$ get the asymptotic behaviour

$$|u|^{1-\beta} |x|^{\beta-1} \beta \exp(-|u|^{2-2\beta} |x|^{2\beta}).$$

We want $q'(x; \theta_q)$ continuously differentiable in order to get $l(x; \theta_L)$ continuously differentiable. Therefore, we propose the following function

$$(16) \qquad q(x; \theta_q) = \begin{cases} x & x < u - \varepsilon \\ \frac{c(x-u-\varepsilon)^{k_1}}{k_1(k_1-1)} + \frac{d(x-u-\varepsilon)^{k_2}}{k_2(k_2-1)} + x & u - \varepsilon \leq x < u + \varepsilon \\ u(x/u)^{\beta} & u + \varepsilon \leq x \end{cases}$$

where $\varepsilon$ is a fixed constant determining the length of the transition zone and $c, d, k_1$ and $k_2$ are chosen in order to get $l(x; \theta_L)$ smooth. We have the following equations

$$(17) \qquad d = \frac{f'(u+\varepsilon) - 1 - \frac{2\varepsilon}{k_1-1} f''(u+\varepsilon)}{(\frac{1}{k_2-1} - \frac{1}{k_1-1})(2\varepsilon)^{k_2-1}}$$

$$(18) \qquad c = \frac{f''(u+\varepsilon) - d(2\varepsilon)^{k_2-2}}{(2\varepsilon)^{k_1-2}}$$

in order to get $q(x; \theta_q)$ twice continuously differentiable where the constants satisfy $k_1 > 3$ and $k_2 > 2$. The function $q$ is smooth with $k_1 = 4$ and $k_2 = 3$. See Figure 4 for an illustration of a $q(x; \theta_q)$ function and the corresponding density. The parameters in the model, $\theta_G, \beta$ and $u$ should be found from data. The length of the mixing zone should be set as a constant or connected to the variance of $G(x)$ since it is difficult to estimate this from data. Similarly to combine several components in (5), we may have several mixing zones in (16).

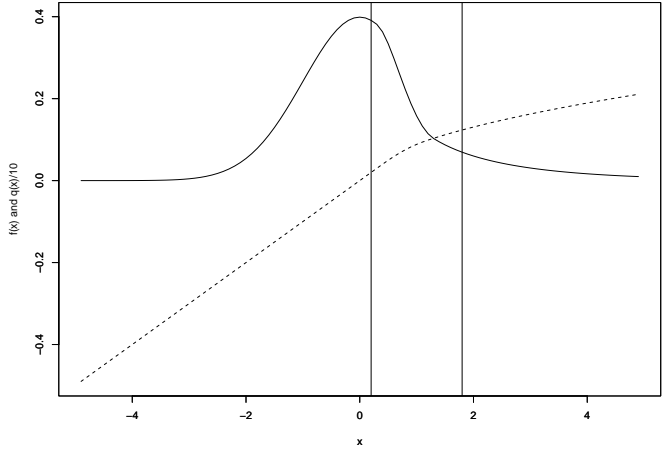

FIGURE 4. The density in the transformed normal model with $g(x; \theta_G) \sim N(0, 1)$. The figure also shows $q(x, \theta_q)/10$ with mixing zone $(0.3, 1.3)$ and $q(x, \theta_q) = (x/u)^{0.5}$

## 4. COMPARISON WITH OTHER MODELS

The traditional method in extreme value modelling is to fix one or two thresholds, and use only values further out in the tails than these thresholds for parameter estimation. By using Equation (5) with three components, fixing the $u_i$'s in advance, and letting $\varepsilon_i = 0$ for all components, the cdf-method proposed is identical with the traditional one.

Equation (1) bears resemblance with the mixed model

$$(19) \qquad l_2(x; \theta_l) = \frac{1}{Z(\theta_l)}(p(x; \theta_p)g(x; \theta_G) + (1 - p(x; \theta_p))f(x; \theta_F)),$$

proposed by [11]. Here $f$ and $g$ are the densities of $F$ and $G$ respectively, and $Z(\theta_l)$ is an integrational constant. The integrational constant is generally found by numerical integration, which is likely to make the maximum likelihood estimation unstable and computationally expensive. By mixing the cdfs instead of the densities, we often get analytic expressions for the integrational constant, and the parameter estimation becomes more stable. Otherwise, it makes little difference whether the mixing is based on the densities or the cdfs. However, the increased efficiency of our model as compared to Equation (19) makes it more manageable to use in several dimensions.

In [3] it is proposed to use the cdf

$$(20) \qquad L(x; \theta_L) = \begin{cases} G(x; \theta_G) & x < u \\ G(u; \theta_G) + (1 - G(u; \theta_G))F(x; \theta_F) & x \geq u \end{cases}$$

This is identical with the mixing model (1) if we assume there is no mixing zone, i.e. $\varepsilon = 0$ and $f(x)$ is replaced with $cf(x)$ for a constant $c$ such that $\kappa = 1$. By introducing a mixing zone we obtain a continuous density. As shown in the example, this does not imply an increase in the number of parameters that need to be estimated. It only makes the result more plausible and offers more stable estimation of the parameters since the density is smooth.

In the recent preprint [5] another model of the same type is proposed in the context of neural networks. It is proposed to mix a normal distribution with a GPD distribution with the restriction on the parameters such that the density and the derivative of the density are the same on both sides of the thresholds. This gives a smooth density without a mixing zone. Their model has one parameter less than the GPD-normal model presented in this paper since the requirement of a continuous derivative of the density eliminates the scaling parameter in the GPD density. This implies that variance of the normal distribution is connected to the scaling of the tail in the mixed model. It is neither easy to generalize their model to other densities than GPD nor to several dimensions.

## 5. NUMERICAL APPLICATIONS

We illustrate the proposed models on both synthetic data and real financial data. All parameter estimation is performed by maximizing the log-likelihood. The maximisation is done numerically using the routine *nlminb* in R. This seems to work very well in all tests performed.

5.1. **Synthetic data.** In this section, we test three different models. We generate synthetic data from one of the models and then estimate parameters and quantiles in all the proposed models. The first two models are mixtures of the form defined by Equation (5) with $k = 3$ and the normal distribution as the central distribution. The first model has the generalized Pareto Distribution (GPD) distribution in both tails and the second has the Weibull distribution in both tails. The GPD cdf is

$$G(x; \xi, \sigma) = 1 - (1 + \frac{\xi x}{\sigma})^{-\frac{1}{\xi}}$$

assuming $\xi > 0$, $\sigma > 0$ and $x > 0$, and the cdf in the Weibull distribution is

$$G(x; \beta, \lambda) = 1 - \exp(-(x\lambda)^\beta)$$

for $\beta > 0$, $\lambda > 0$ and $x > 0$. This gives 10 parameters in the model described by Equation (5), 2 in each of the three distributions in addition to $u_1, u_2, \varepsilon_1$ and $\varepsilon_2$.

TABLE 1. Average of estimated parameters and standard deviation of the estimates when simulated using a GPD-Normal-GPD distribution. $10m$ indicates the use of 10.000 data points in the sample.

| | $\theta_1$ | $\theta_2$ | $\sigma_2$ | $\theta_4$ | $\theta_5$ | $u_1$ | $u_2$ |
|---|---|---|---|---|---|---|---|
| Simulation GPD-N-GPD | 0.300 | 0.400 | 1.000 | 0.200 | 0.400 | -2.166 | 2.415 |
| Est. GPD-N-GPD $10m$ | 0.299 | 0.401 | 1.000 | 0.200 | 0.400 | -2.165 | 2.417 |
| St.dev. GPD-N-GPD $10m$ | 0.019 | 0.017 | 0.010 | 0.021 | 0.020 | 0.067 | 0.081 |
| Est. GPD-N-GPD | 0.295 | 0.404 | 0.996 | 0.195 | 0.403 | -2.131 | 2.390 |
| St. dev. GPD-N-GPD | 0.068 | 0.058 | 0.035 | 0.063 | 0.061 | 0.260 | 0.311 |
| Est. Weibull-N-Weibull | 0.511 | 0.211 | 1.000 | 0.606 | 0.253 | -2.391 | 2.513 |
| St.dev. Weibull-N-Weibull | 0.058 | 0.061 | 0.031 | 0.061 | 0.059 | 0.170 | 0.315 |
| Est. transf. N | - | 0.410 | 1.060 | - | 0.489 | -1.755 | 1.914 |
| St.dev. transf. N | - | 0.069 | 0.031 | - | 0.104 | 0.190 | 0.251 |

The thresholds $u_1$ and $u_2$ are determined from Equation (12). This reduces the number of parameters and also gives smoother distributions. The length of the transition intervals, $2\varepsilon_i$, are not critical in the estimation and not easy to estimate. Hence, we set $\varepsilon_i = \sigma_2$ for $i = 1$ and $i = 2$ where $\sigma_2$ is the standard deviation of the central normal distribution. In all the tests we set the expectation in the central density equal to 0 leaving 5 unknown parameters.

The third model we test is the transformation model described in Equation (14) with the polynomial function for $q$ given in Equation (16). Also here we set $\varepsilon_i = \sigma_2$ for $i = 1$ and $i = 2$ in order to make the same choice as in the previous model. We denote $\sigma_2$ as the standard deviation of the normal distribution in order to use the same symbol with the corresponding parameter in the other models. Also here there are 5 unknown parameters.

In the tests we simulate $m = 1000$ samples from each of the models in turn and then estimate parameters in all the models. In addition, we also test with $10m = 10.000$ samples with the same model as is used in the simulation. This is repeated $k = 500$ where we estimate the parameters in each of the three models, the corresponding $0.001, 0.01, 0.99$ and $0.999$ quantiles, the difference in $L_1$ norm, and the log-likelihood value. The tables give the average and the standard deviation of the estimated parameters/quantiles/values. The difference in $L_1$ norm is estimated by dividing the state space into 100 intervals. The difference in $L_1$ norm is half the sum of the absolute value of the difference in probability between the estimated and the original density in each interval. Simulation from a distribution where the density differs in the $L_1$ norm by 0.01 compared to the correct density, implies that 0.01 of the samples are from a wrong distribution.

The parameters are estimated by maximizing the log-likelihood. The model with GPD distributions in the tails is left with the following 5 parameters: $\xi_1, \sigma_1, \sigma_2, \xi_3$, $\sigma_3$ and the model with Weibull distribution in the tails has the parameters $\beta_1, \lambda_1, \sigma_2$, $\beta_3, \lambda_3$. In these two models the thresholds $u_1$ and $u_2$ are found from Equation (12) based on the other parameters. The model with transformations has the parameters $u_1, u_2, \beta_1, \sigma_2, \beta_3$. In Tables 1, 3, and 5, the parameters are denoted $\theta_1, \theta_2, \sigma_2, \theta_3, \theta_4$, $u_1$, and $u_2$ where $\theta_1, \theta_2, \theta_3, \theta_4$, have a different interpretation in the different models. The results of the simulations are shown in Tables 1 - 6. We see that there are quite good estimates for all the parameters. The standard deviation is comparable with estimation of $\sigma$ in the normal distribution with the same sample size. Only 1-4% of the $m = 1000$ samples are from the tails and in the mixing zone these are mixed with data points from the central distribution. The standard deviations for the different parameters have about the same size including $\sigma_2$, the standard deviation in the central distribution. The estimates for the thresholds $u_i$ have a larger standard deviation than the other parameters, indicating that the threshold should be determined implicitly by the other parameters. We have also tried to estimate these simultaneously with the other parameters. Then all parameters have larger uncertainty. Also the quantiles and the density measured in the $L_1$ error are quite close to the quantiles and density that were used in the simulation.

TABLE 2. Quantiles, $L_1$-error and log-likelihood using GPD-N-GPD.

|  | $q_{0.001}$ | $q_{0.01}$ | $q_{0.99}$ | $q_{0.999}$ | $L_1$ | loglikeh. |
|---|---|---|---|---|---|---|
| Simulation GPD-N-GPD | -9.15 | -3.92 | 3.00 | 5.91 | - | -1583.5 |
| Est. GPD-N-GPD 10$m$ | -9.14 | -3.92 | 3.00 | 5.91 | 0.005 | -15833.0 |
| St.dev. GPD-N-GPD 10$m$ | 0.54 | 0.12 | 0.08 | 0.31 | 0.002 | 99.5 |
| Est. GPD-N-GPD | -9.20 | -3.90 | 2.99 | 5.88 | 0.016 | -1581.2 |
| St. dev. GPD-N-GPD | 1.96 | 0.41 | 0.24 | 0.96 | 0.007 | 30.5 |
| Est. Weibull-N-Weibull | -9.34 | -4.12 | 3.10 | 6.15 | 0.015 | -1583.6 |
| St.dev. Weibull-N-Weibull | 1.70 | 0.49 | 0.28 | 0.85 | 0.008 | 33.2 |
| Est. transf. N | -8.41 | -4.08 | 3.22 | 5.95 | 0.025 | -1583.8 |
| St.dev. transf. N | 1.66 | 0.50 | 0.32 | 1.01 | 0.006 | 33.4 |

TABLE 3. Average of estimated parameters and standard deviation of the estimates when simulated using a Weibull-Normal-Weibull distribution.

|  | $\theta_1$ | $\theta_2$ | $\sigma_2$ | $\theta_4$ | $\theta_5$ | $u_1$ | $u_2$ |
|---|---|---|---|---|---|---|---|
| Simulation Weibull-N-Weibull | 0.500 | 0.200 | 1.000 | 0.600 | 0.250 | -2.394 | 2.487 |
| Est. Weibull-N-Weibull 10$m$ | 0.502 | 0.202 | 0.999 | 0.600 | 0.251 | -2.388 | 2.483 |
| St.dev. Weibull-N-Weibull 10$m$ | 0.019 | 0.021 | 0.010 | 0.021 | 0.021 | 0.052 | 0.063 |
| Est. Weibull-N-Weibull | 0.511 | 0.211 | 1.000 | 0.606 | 0.253 | -2.390 | 2.514 |
| St.dev. Weibull-N-Weibull | 0.058 | 0.061 | 0.031 | 0.061 | 0.059 | 0.170 | 0.315 |
| Est. GPD-N-GPD | 0.383 | 0.331 | 1.000 | 0.333 | 0.298 | -2.314 | 2.590 |
| St.dev. GPD-N-GPD | 0.091 | 0.071 | 0.035 | 0.125 | 0.111 | 0.265 | 0.440 |
| Est. transf. N | - | 0.410 | 1.060 | - | 0.489 | 1.755 | 1.914 |
| St.dev. transf. N | - | 0.069 | 0.031 | - | 0.104 | 0.190 | 0.251 |

TABLE 4. Quantiles, $L_1$-error and log-likelihood using Weibull-Normal-Weibull.

|  | $q_{0.001}$ | $q_{0.01}$ | $q_{0.99}$ | $q_{0.999}$ | $L_1$ | loglikeh. |
|---|---|---|---|---|---|---|
| Simulation Weibull-N-Weibull | -9.45 | -4.18 | 3.12 | 6.21 | - | -1585.7 |
| Est. Weibull-N-Weibull 10$m$ | -9.45 | -4.18 | 3.13 | 6.22 | 0.005 | - 15874.9 |
| St.dev. Weibull-N-Weibull 10$m$ | 0.51 | 0.14 | 0.08 | 0.28 | 0.002 | 99.4 |
| Est. Weibull-N-Weibull | -9.34 | -4.12 | 3.10 | 6.15 | 0.015 | -1583.6 |
| St.dev. Weibull-N-Weibull | 1.70 | 0.49 | 0.278 | 0.848 | 0.008 | 33.2 |
| Est. GPD-N-GPD | -11.41 | -4.08 | 3.11 | 7.82 | 0.017 | -1584.7 |
| St.dev. GPD-N-GPD | 3.02 | 0.51 | 0.34 | 2.18 | 0.008 | 33.4 |
| Est. transf. N | -8.41 | -4.08 | 3.22 | 5.95 | 0.025 | -1583.8 |
| St.dev. transf. N | 1.66 | 0.50 | 0.32 | 1.01 | 0.006 | 33.4 |

TABLE 5. Average of estimated parameters and standard deviation of the estimates when simulated using a transformed normal distribution. The transformed distribution uses $u_i$ as one of the five estimated parameters (instead of $\theta_1$ and $\theta_4$) while $u_i$ depend on the other parameters in the other models.

|  | $\theta_1$ | $\theta_2$ | $\sigma_2$ | $\theta_4$ | $\theta_5$ | $u_1$ | $u_2$ |
|---|---|---|---|---|---|---|---|
| Simulation transf. N | - | 0.450 | 1.000 | - | 0.600 | -1.500 | 1.500 |
| Est. transf. N 10$m$ | - | 0.450 | 1.000 | - | 0.600 | -1.498 | 1.500 |
| St.dev. transf. N 10$m$ | - | 0.015 | 0.008 | - | 0.021 | 0.034 | 0.052 |
| Est. transf. N | - | 0.458 | 0.996 | - | 0.605 | 1.484 | 1.489 |
| St.dev. transf. N | - | 0.056 | 0.028 | - | 0.075 | 0.118 | 0.198 |
| Est. GPD-N-GPD | 0.343 | 0.360 | 0.928 | 0.241 | 0.387 | -1.985 | 2.001 |
| St.dev. GPD-N-GPD | 0.075 | 0.056 | 0.038 | 0.116 | 0.106 | 0.221 | 0.467 |
| Est. Weibull-N-Weibull | 0.554 | 0.256 | 0.926 | 0.685 | 0.339 | -2.074 | 2.028 |
| St.dev. Weibull-N-Weibull | 0.055 | 0.059 | 0.033 | 0.078 | 0.080 | 0.159 | 0.298 |

TABLE 6. Quantiles, $L_1$-error and log-likelihood using a transformed normal distribution.

|  | $q_{0.001}$ | $q_{0.01}$ | $q_{0.99}$ | $q_{0.999}$ | $L_1$ | loglikeh. |
|---|---|---|---|---|---|---|
| Simulation transf. N | -7.51 | -4.00 | 3.13 | 5.02 | - | -1548.2 |
| Est. transf. N 10$m$ | -7.51 | -4.00 | 3.13 | 5.02 | 0.005 | -15483.9 |
| St.dev. transf. N 10$m$ | 0.33 | 0.11 | 0.07 | 0.17 | 0.002 | 95.8 |
| Est. transf. N | -7.52 | -3.99 | 3.11 | 5.02 | 0.018 | -1545.7 |
| St.dev. transf. N | 1.17 | 0.38 | 0.20 | 0.54 | 0.008 | 31.1 |
| Est. GPD-N-GPD | -10.2 | -3.94 | 3.12 | 6.79 | 0.028 | - 1550.2 |
| St.dev. GPD-N-GPD | 2.31 | 0.42 | 0.29 | 1.76 | 0.007 | 31.5 |
| Est. Weibull-N-Weibull | -8.44 | -3.98 | 3.10 | 5.71 | 0.025 | 1548.0 |
| St.dev. Weibull-N-Weibull | 1.27 | 0.40 | 0.24 | 0.63 | 0.006 | 31.3 |

The uncertainty in the $q_{0.001}$ and $q_{0.999}$ quantiles in the GPD distribution is larger than for the other distributions since this has heavier tails than the two other distributions. As expected, we always get better estimates when we fit the same model as is used in the simulation and when we increase the number of samples to $10m = 10.000$.

5.2. **Financial data.** We want to illustrate the use of the models on real data and have selected three stock market indices; the European, the American and the Japanese. It is not our ambition to suggest the best possible model for these data. That would require a more complex model including for example handling of stochastic volatility which is outside the topic of this paper. We study each of the stock markets independently using the methods from Sections 2 and 3.

We assume that the three stock markets can be represented by the corresponding Morgan Stanley (MSCI) price indices in local currency neglecting the currency risk in the portfolio. We use index data from the period 01.01.1987 to 28.05.2002 for model estimation. This period corresponds to $m = 4065$ observations. The return series are shown in Figure 5. Let $x_{i,t}$ denote the original indices, $i = 1, 2, 3$. We use the data $r_{i,t} = log(x_{i+1,t}/x_{i,t}) - \mu_i$ where $\mu_i$ is determined such that $\sum_i r_{i,t} = 0$.

Figure 6 shows normal QQ-plots for the standardised logarithmic residuals $r_{i,t}/\sigma_i$ for each of the three markets where $\sigma_i$ is the standard deviation of $r_{i,t}$. As can be seen from the figure, all distributions are doubly heavy-tailed. Moreover, they are clearly skewed, having one tail heavier than the other. This motivates for the use of a mixture distribution with three components, one for the left tail, one for the centre of the distribution, and one for the right tail, respectively. Hence, we use the distribution given in Equation (5) with three components. Exactly as in Section 5.1 we test with the GPD and the Weibull density in both tails and with the normal distribution in the centre. In addition, we test with the transformed normal distribution given in Equation (14).

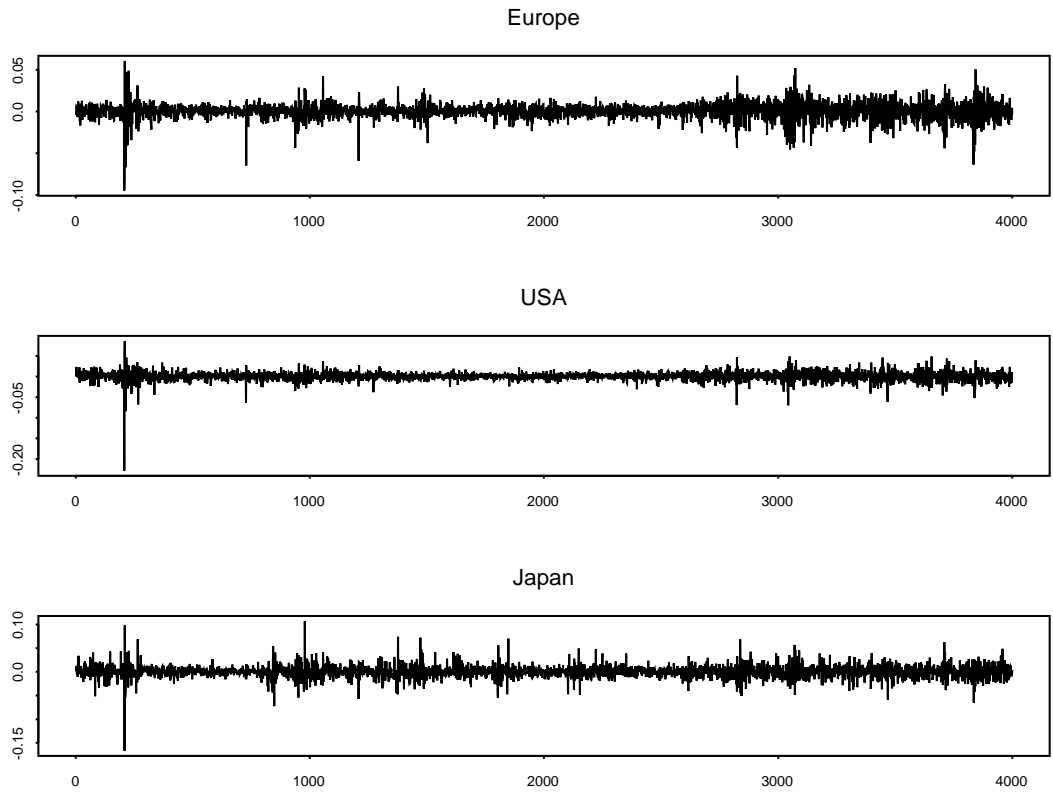

FIGURE 5. European, American and Japanese geometric return
series for the period 01.01.1987 − 28.05.2002.

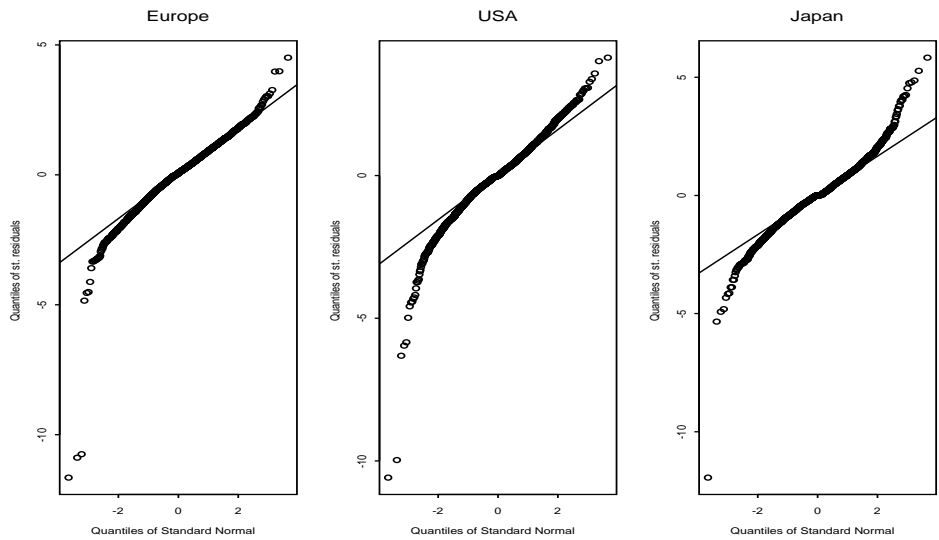

FIGURE 6. QQ-plots of the standardized residuals fitted against
the normal distribution.

TABLE 7. Parameter estimates for the GPD-N-GPD mixture model using residuals. We have $\varepsilon_i = \sigma_2$.

| Parameter | Europe | USA | Japan |
|---|---|---|---|
| $\xi_1$ | 0.266 | 0.156 | 0.432 |
| $\sigma_1$ | 0.00395 | 0.00564 | 0.00831 |
| $\sigma_2$ | 0.00409 | 0.00320 | 0.00377 |
| $\xi_3$ | 0.0735 | 0.198 | 0.0724 |
| $\sigma_3$ | 0.00498 | 0.00680 | 0.00801 |
| log-likelihood | 13609 | 13227 | 12339 |

TABLE 8. Parameter estimates for the Weibull-N-Weibull mixture model using residuals. We have $\varepsilon_i = \sigma_2$.

| Parameter | Europe | USA | Japan |
|---|---|---|---|
| $\beta_1$ | 0.686 | 0.801 | 0.987 |
| $\lambda_1$ | 0.00357 | 0.00546 | 0.00873 |
| $\sigma_2$ | 0.00454 | 0.00344 | 0.00368 |
| $\beta_3$ | 0.885 | 0.977 | 0.948 |
| $\lambda_3$ | 0.00464 | 0.00664 | 0.00845 |
| log-likelihood | 13612 | 13223 | 12334 |

In all cases we have the same 5 parameters as described in Section 5.1 that are estimated by maximizing the likelihood

$$(21) \qquad \prod_{t=1}^{m} l(r_{i,t}; \theta_L),$$

where $l(r; \theta_L)$ is given by Equations (6) and (15) respectively. The results are shown for each of the models in Tables 7 - 9. For the transformed normal model, three of the threshold values ended up equal to the limit $\pm 0.005$. This indicates that we get best fit with using the transformation for all negative/positive values. The parameter $\sigma_2$ is then only used to set the density for $x = 0$ and influences the joint density in the mixing zones $(-0.01, 0)$ and $(0, 0.01)$. We get best fit using GPD-N-GPD, then Weibull-N-Weibull and then transformed normal density. The estimated quantiles are given in Table 11 together with the corresponding results from the multivariate distributions.

5.2.1. *Parameter estimates for NIG distribution.* We have chosen to compare the results using the mixture model with the ones obtained using the Normal Inverse Gaussian (NIG) distribution. This distribution has been used for financial applications, both as the conditional distribution of a GARCH-model, see [2], and as the unconditional return distribution, see [4]. The paper [19] compares different probability distributions for the innovations in one-dimensional processes. The authors consider a NIG distribution, a skewed Student's t-distribution and a non-parametric kernel approximation. They report that the NIG distribution provides the best fit overall for the models considered.

TABLE 9. Parameter estimates for the transformed normal model using residuals. We have $\varepsilon_i = 0.005$.

| Parameter | Europe | USA | Japan |
|---|---|---|---|
| $u_1$ | -0.007 | -0.006 | - 0.005 |
| $u_2$ | 0.0085 | 0.005 | 0.005 |
| $\beta_1$ | 0.550 | 0.565 | 0.665 |
| $\sigma_2$ | 0.00674 | 0.00659 | 0.00778 |
| $\beta_3$ | 0.583 | 0.665 | 0.655 |
| log-likelihood | 13591 | 13188 | 12300 |

TABLE 10. Parameter estimates for NIG distributions for logarithmic residuals.

| Parameter | Europe | USA | Japan |
|---|---|---|---|
| $\mu$ | 0.00134 | 0.000746 | -0.000308 |
| $\delta$ | 0.00678 | 0.00733 | 0.00945 |
| $\alpha$ | 78.3 | 67.2 | 57.0 |
| $\beta$ | -12.6 | -3.85 | 1.17 |

The normal inverse Gaussian (NIG) distribution is a generalised hyperbolic distribution with $\lambda = -\frac{1}{2}$. Its density is

$$f_x(x) = \frac{\delta \, \alpha \, \exp\left(\delta\sqrt{\alpha^2 - \beta^2}\right) \, K_1\left(\alpha \sqrt{\delta^2 + (x-\mu)^2}\right) \exp\left(\beta \, (x-\mu)\right)}{\pi \, \sqrt{\delta^2 + (x-\mu)^2}},$$

where $\delta > 0$ and $0 < |\beta| \leq \alpha$. In the above expression, $K_1$ is the modified Bessel function of the third kind of order 1, see [1]. The parameters $\mu$ and $\delta$ determine the location and scale, respectively, while $\alpha$ and $\beta$ control the shape of the density. In particular, $\beta = 0$ corresponds to a symmetric distribution.

The parameters of the NIG distribution are estimated using the EM-algorithm described in [13], with the moment estimates as starting values. The parameter estimates are shown in Table 10.

5.2.2. *Comparing the models in the tails.* We have used graphical logarithmic left and right hand tail tests to examine the fit in the tails. The graphical tests were performed as follows. Let $(X_{(1)}, ..., X_{(N)})$ denote the order statistic of the historical data, and $\hat{F}(x)$ the estimated cumulative distribution function of the fitted distribution. For the NIG distribution this is calculated using the method described in [15]. A plot of $\log(\hat{F}(X_{(t)}))$ against $X_{(t)}$ superimposed on a plot of $\log(t/(N+1))$ against $X_{(t)}$ shows the left tail fit for the fitted distribution, and a plot of $\log(1 - \hat{F}(X_{(t)}))$ against $X_{(t)}$, superimposed on a plot of $\log((N+1-t)/(N+1))$, the right tail fit.

Figure 7 shows the plots. All the models give quite similar results but the mixture distribution with GPD tails followed by the NIG distribution seems slightly better than the others.

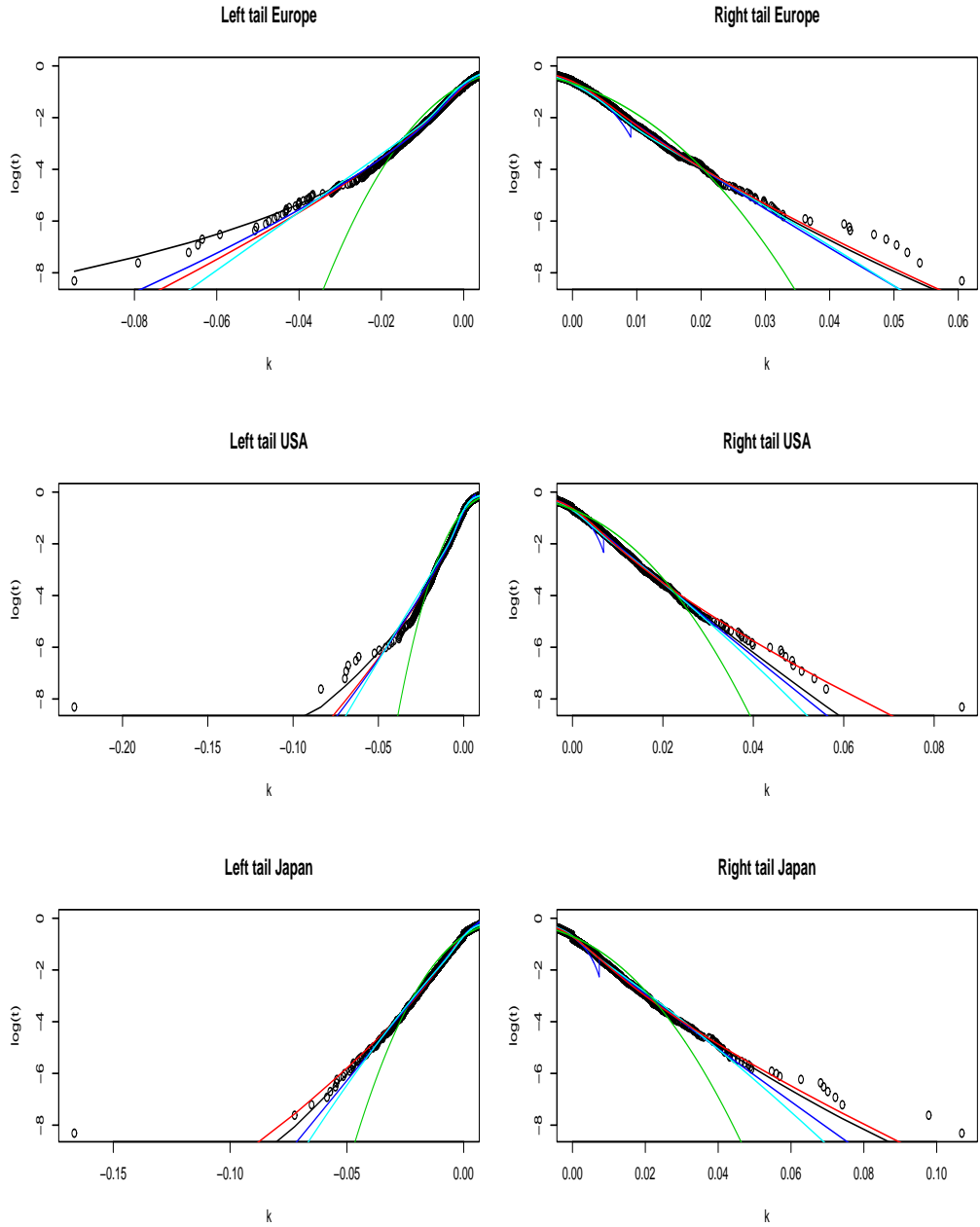

FIGURE 7. Plot of the tail behaviour in the five models. The circles correspond to the empirical data, the light-blue line to the mixture distribution with Weibull tails, the black line to the mixture distribution with GPD tail, the red to the NIG distribution and the blue line to the transformed normal. For reference, a normal fit is also included shown in green.

TABLE 11. Comparing the quantiles and likelihood between the different models and the data. The first four lines are the empirical data and then the models using GPD, Weibull and transformation that assumes independence between the three variables. NOP is number of parameters, $q_{\alpha,p}$ the quantiles and loglik. the log-likelihood of the estimated variables.

| Model | NOP | p | $q_{E,p}$ | $q_{U,p}$ | $q_{J,p}$ | $q_{A,p}$ | loglik. |
|-------|-----|---|-----------|-----------|-----------|-----------|---------|
| Data  |     | 0.01 | -0.0293 | -0.0273 | -0.0345 | -0.0114 | |
|       |     | 0.99 | 0.0235 | 0.0272 | 0.0359 | 0.0110 | |
|       |     | 0.001 | -0.0645 | -0.0688 | -0.0585 | -0.0171 | |
|       |     | 0.999 | 0.0504 | 0.0506 | 0.0721 | 0.0285 | |
| U.GPD-N-GPD | 3x5 | 0.01 | -0.0302 | -0.0310 | -0.0357 | - 0.00580 | 39178 |
|       |     | 0.99 | 0.0244 | 0.0280 | 0.0365 | 0.00626 | |
|       |     | 0.001 | -0.0683 | -0.0600 | -0.0595 | -0.0113 | |
|       |     | 0.999 | 0.0413 | 0.0454 | 0.0632 | -0.0115 | |
| U. W-N-W | 3x5 | 0.01 | -0.0299 | -0.0311 | -0.0349 | -0.000569 | 39168 |
|       |     | 0.99 | 0.0241 | 0.0276 | 0.0358 | 0.00511 | |
|       |     | 0.001 | -0.0558 | -0.0547 | -0.0557 | -0.0115 | |
|       |     | 0.999 | 0.0391 | 0.0439 | 0.0582 | 0.0117 | |
| U. NT | 3x5 | 0.01 | -0.0306 | -0.0325 | -0.0350 | -0.00875 | 39079 |
|       |     | 0.99 | 0.0244 | 0.0273 | 0.0360 | 0.00895 | |
|       |     | 0.001 | -0.0512 | -0.0535 | -0.0535 | -0.0105 | |
|       |     | 0.999 | 0.0397 | 0.0418 | 0.0553 | 0.0111 | |

## 6. SUMMARY AND CONCLUSIONS

In this paper we present a new method to mix densities from different models. The method is inspired by [11]. But we mix cdfs instead of densities since this is much more computationally stable and efficient making it possible to extend to several dimensions in contrast to the method described in [11]. The paper [3] also combines cdfs but by introducing a mixing zone we obtain continuous densities. We also show how univariate transformations may be used in order to represent tail behaviour.

The different models are tested by simulation from one model and then estimate parameters and quantiles from all the models. The suitability of each model depends on the data in each case. We compare the different models on a financial data set, evaluating likelihood and tail behaviours. The different models seem to behave quite similar.

Before we select a model we should analyse the data and then find a model with suitable tail behaviour. In the multivariate case we need to find the univariate tail behaviour and the correlation in the tails of the data.

## ACKNOWLEDGMENTS

This work is partly sponsored by the Norwegian fund Finansmarkedsfondet and partly by the Norwegian Research Council. We thank Kjersti Aas and Håvard Rue for valuable comments. We also thank the anonymous referees and assistant editor for many valuable comments.

Norwegian Computing Center, Oslo, Norway

*E-mail address*: Lars.Holden@nr.no